\documentclass[prb,twocolumn,aps,showpacs,superscriptaddress]{revtex4}

\usepackage{amsfonts}
\usepackage{graphicx}
\usepackage{textcomp}
\usepackage{txfonts}
\usepackage{bm}
\usepackage{CJK}

\begin{document}

\title{Enhanced Josephson tunneling between high temperature superconductors through a normal pseudogap underdoped cuprate with a finite energy cooperon
}

\author{Kun Huang}
\affiliation{Department of Physics, and Center of Theoretical and Computational Physics, The University of Hong Kong,
 Hong Kong, China}
\author{Wei-Qiang Chen}
\affiliation{Department of Physics, South University of Science and Technology of China, Shenzhen, Guangdong, China}
\affiliation{Department of Physics, and Center of Theoretical and Computational Physics, The University of Hong Kong,
 Hong Kong, China}
\author{T. M. Rice}
\affiliation{Institut f\"{u}r Theoretische Physik, ETH Z\"{u}rich,CH-8093 Z\"{u}rich, Switzerland}
\affiliation{Department of Physics, and Center of Theoretical and Computational Physics, The University of Hong Kong,
 Hong Kong, China}
\affiliation{Condensed Matter Physics and Materials Science Department,Brookhaven National Laboratory, Upton, NY 11973, USA}
\author{F. C. Zhang}
\affiliation{Department of Physics, and Center of Theoretical and Computational Physics, The University of Hong Kong,
 Hong Kong, China}

\date{\today}

\begin{abstract}
  The Josephson coupling between optimally cuprate superconductors separated by a spacer with a finite energy cooperon
  excitation which contributes to the Josephson coupling strength, is examined.  For an underdoped cuprate barrier in its normal state, the YRZ model gives a good
  description of the temperature dependent enhanced Josephson coupling.  A detailed examination of origin of the
  enhancement shows a significant contribution from the cooperon excitation which is comparable to that from nodal quasiparticles.
\end{abstract}


\maketitle

There have been a number of reports of enhanced, even giant, penetration lengths for d-wave
superconductivity into nonsuperconducting phases of underdoped cuprates \cite{Triscone-JLTP-1996,Bozovic-PRL-2004,Bozovic-Meissner-effect,Wojek-PRB-2012}.
Bozovic et al. reported giant proximity effect in uniform trilayer junctions of high temperature superconductors at temperature $T$ above the superconducting transition temperature $T_c$ of the spacer underdoped cuprate layer in the pseudogap phase\cite{Bozovic-PRL-2004}. The enhanced penetration length observed in this case has been explained by a model with strong phase fluctuations due to a reduced
phase stiffness at underdoping\cite{Marchand-fluctuating}.

Recently it has been proposed
that the opening of the pseudogap in the antinodal regions of $k$-space in underdoped cuprates,
is accompanied by the appearance of a finite energy bound hole pair excitation, also known as a cooperon\cite{Rice-four-legs,Kaiyu-Troyer}. In this letter we examine the influence of
such cooperon excitations on the penetration length of superconducting order into nonsuperconducting cuprate materials.  A cooperon  creates a low energy pole in the Cooper pair correlation function leading to enhanced penetration of  superconducting order.  We show that the cooperon contribution to the enhancement of the Josephson tunneling in thick layer is significant.

We are motivated by the recent ARPES\cite{YRZ,Yang-nature-08,Yang-EPL-09,Yang-PRL-11}
(angle resolved photoemission spectroscopy) and other experiments\cite{AIPES-PRB-09,Kohsaka-Nature-08,
LeBlanc-PRB-specificheat,Illes-PRB-optical,Carbotte-PRB-Penetration,LeBlanc-PRB-Raman,
Leni-PRL-07,Meng-Nature-09,Kaiyu-Andreev}, which support rather a 2-gap scenario with an insulating pseudogap for underdoped cuprates. This leads us to examine the superconducting penetration at $T>T_c$ , using the phenomenological 2-gap YRZ theory put forward by Yang et al. for the pseudogap phase. The YRZ model has had considerable success in describing many anomalous properties of the pseudogap phase. For a recent review see Rice,Yang and Zhang\cite{Rice-Review}. As discussed below, this model has a low energy cooperon pole at temperatures $T > T_c$ which enhances the penetration.

  An analysis of the single particle propagator in a 2-dimensional array of 2-leg Hubbard ladders was an important input in the formulation of the YRZ ansatz. A recent extension by Konik, Rice and Tsvelik to an array of 4-leg  Hubbard ladders with 4 bands crossing the Fermi surface. They showed that at low doping there is an insulating gap with a cooperon resonance in the outer band pair which coexists with metallic behavior in the inner band pair. The virtual exchange of the cooperon resonance with support on the outer bands introduces a pairing mechanism in the inner band Fermi surface\cite{Rice-four-legs}.

  A decade ago Nozieres and Pistolesi\cite{Nozieres-EPJ-99} studied the transition for fermions with an attractive interaction between a superconducting and a semiconducting state within  BCS mean field theory as the one particle band gap is increased. They found that at a critical value of the band gap, which is simply related to the superconducting gap of the starting metallic state, the pairing amplitude and phase stiffness dropped continuously to zero while the single particle gap remained finite. Recently Rice \textit{et al.} \cite{Rice-Review} showed that the approach to this quantum critical point(QCP)  from the semiconducting state was characterized by a softening of the cooperon mode, dropping to zero energy at the QCP\cite{Rice-Review}.
     Support for the existence of cooperons in strongly underdoped cuprates also comes from exact diagonalization results of the strong coupling t-J models on small clusters. When extrapolated to an infinite lattice, these show a low lying cooperon resonance in the antinodal regions of the Brillouin zone\cite{Rice-four-legs}.

   We use the phenomenological YRZ theory to describe the pseudogap phase with input parameters chosen to have the same values as in the original work\cite{YRZ} to model the experiments by Bozovic et. al. on the enhanced superconducting penetration \cite{Bozovic-PRL-2004} into underdoped  LSCO cuprates at temperatures $T > T_c$. In the YRZ model, the coherent part of the Green's function is given by,
   \begin{equation}
   G_{\rm YRZ}({\bf k},\omega)={g_t\over\omega-\varepsilon_{\bf k}-\Sigma_R({\bf k},\omega)}
   \end{equation}
   The self-energy is given by $\Sigma_R({\bf k},\omega)={|\Delta_R({\bf k})|^2/[\omega+\epsilon({\bf k})]}$ with
    $\epsilon_{\bf k}=-2t(\cos k_x+\cos k_y)$, $\varepsilon_{\bf k}=\epsilon_{\bf k}-4t'\cos k_x\cos k_y-2t''(\cos 2 k_x+\cos 2k_y)-\mu_p$, and the RVB gap $\Delta_R({\bf k})=\Delta_0(x)(\cos k_x-\cos k_y)$. $g_t=2x/(1+x)$ is the doping-dependent renormalization factor\cite{RMFT}. $t,~t',~t''$ are the renormalized hopping integrals, and $\mu_p$ is an effective chemical potential determined by Luttinger sum rule that the area enclosed by the four Fermi pockets equals to the doped hole density \cite{YRZ,LSR}.
   The modified pairing form for the self-energy leads to two quasiparticle bands with energy dispersion and the corresponding spectral weight,
   \begin{eqnarray}
   E_{\bf k}^\pm&=&(\varepsilon_{\bf k}-\epsilon_{\bf k})/2\pm\sqrt{(\varepsilon_{\bf k}+\epsilon_{\bf k})^2/4+\Delta_R^2({\bf k})}\label{YRZ_dispersion}\nonumber\\
   Z^\pm_{\bf k}&=&[1+(\varepsilon_{\bf k}+\epsilon_{\bf k})/(2E^\pm_{\bf k}-\varepsilon_{\bf k}+\epsilon_{\bf k})]/2
   \end{eqnarray}

\begin{figure}[htbp]
\centerline{\includegraphics[width=0.3\textwidth]{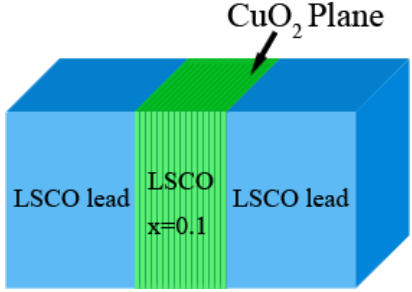}}
\caption[]{\label{fig:setup} Schematic illustration of the setup.  The leads are two optimal doped LSCO material, while
  the center is an underdoped LSCO with doping $x = 0.1$.  The interface is parallel with the $CuO_2$ plane. }
\end{figure}

We introduce superconductivity by adding a pairing potential with d-wave symmetry. The magnitude of the pairing
potential is chosen to reproduce mean-field values for $T_c$ in the Bozovic experiments, i.e. $T_c = 25K$ (doping
$x=0.1$) in the underdoped junction and $T_c = 45K$ (doping $x'=0.15$) in the optimum-doped leads. The tunneling setup
is represented by a set of layers, typically 10 for each of the leads and 10 for the junction, as shown in
fig.~\ref{fig:setup}. Next we obtain a set of coupled Bogoliubov-de Gennes (BdG) equations for the layers,
\begin{equation}
\sum_j H^\alpha_{ij}
\left[ \begin{array}{c}
    u^\alpha_{n{\bf k}_\parallel}(j)\\\\
    v^\alpha_{n{\bf k}_\parallel}(j)
\end{array}\right]
=\mathcal{E}^\alpha_{n{\bf k}_\parallel}
\left[ \begin{array}{c}
    u^\alpha_{n{\bf k}_\parallel}(i)\\\\
    v^\alpha_{n{\bf k}_\parallel}(i)
\end{array}\right]\label{BdG}
\end{equation}
where the BdG Hamiltonian $H^\alpha_{ij}$ is given by,
\begin{equation}H^\alpha_{ij}=
\left[\begin{array}{cc}
E^\alpha_i({\bf k}_\parallel)\delta_{ij}-t_{\bot,ij}({\bf k}_\parallel)~,&\Delta_{i}({\bf k}_\parallel)\delta_{ij}\\\\
\Delta^*_{i}({\bf k}_\parallel)\delta_{ij}~,&-E^\alpha_i({\bf k}_\parallel)\delta_{ij}+t_{\bot,ij}({\bf k}_\parallel)
\end{array}\right]\label{Hamiltonian}
\end{equation}
with $i,j$ the layer index,  ${{\bf k}_\parallel}$ the in-plane momentum, $E^\alpha_i({\bf k}_\parallel)$ the YRZ quasiparticle dispersion given by Eq.(\ref{YRZ_dispersion}) which depends on the doping value of of layer-$i$. $t_{\bot,ij}({\bf k}_\parallel)=t_c(\cos k_x-\cos k_y)^2/4$ is the standard hopping between nearest-neighbor layers\cite{XiangTao-PRL-96}$\langle ij\rangle$, and
$\Delta_{i}({\bf k}_\parallel)=\Delta_i d({\bf k}_\parallel)$ is the superconducting gap of layer-$i$ with the $d$-wave factor $d({\bf k}_\parallel)=\cos k_x-\cos k_y$. $\mathcal{E}^\alpha_{n{\bf k}_\parallel}$ and $[u^\alpha_{n {\bf k}_\parallel},~~v^\alpha_{n {\bf k}_\parallel}]^T$ are the eigenvalues and the corresponding orthonormal eigenvectors of BdG equation.

The Nambu Green's function for this multi-layer system can be constructed by the wave-functions as\cite{BdG-textbook},
\begin{equation}
 \mathcal{G}_{ij}({\bf k},i\omega_\nu)=\sum_{\alpha,n}{Z^\alpha_{{\bf k}_\parallel}\over i\omega_\nu-\mathcal{E}^\alpha_{n{{\bf k}_\parallel}}}
\left[ \begin{array}{c}
    u^\alpha_{n{{\bf k}_\parallel}}(i)\\\\
    v^\alpha_{n{{\bf k}_\parallel}}(i)
\end{array}\right]
\left[u^{\alpha}_{n{{\bf k}_\parallel}}(j)~,~~v^{\alpha}_{n{{\bf k}_\parallel}}(j)\right]^*\label{YRZ}
\end{equation}
 Choosing a $d$-wave attractive interaction $V_i({\bf k}_\parallel,{\bf q}_\parallel)=V_i d({\bf k}_\parallel)d({\bf q}_\parallel)$ with $V_i<0$, then the self-consistent superconducting-gap equation is given by,
\begin{equation}
\Delta_i=|V_i|{\rm Im}\sum_{{\bf k}_\parallel}\int{{\rm d} \omega\over2\pi}d({{\bf k}_\parallel})f(\beta\omega)[F^a_{ii}({{\bf k}_\parallel},\omega)-F^r_{ii}({{\bf k}_\parallel},\omega)]
\label{gap_eqYRZ}\end{equation}
with $\beta=1/(k_BT)$ the reciprocal temperature, $f(x)=1/(e^x+1)$ the Fermi function, $F^{r/a}$ the retarded/advanced anomalous Green's function. According to Eq.(\ref{YRZ}), gap equation Eq.(\ref{gap_eqYRZ}) can be simplified as,
\begin{equation}
\Delta_i
={|V_i|\over2}\sum_{{\bf k}_\parallel}\sum_{n\alpha=\pm} d({\bf k}_\parallel) {f(\beta \mathcal{E}^\alpha_{n{\bf k}_\parallel})u_{n{\bf k}_\parallel}^\alpha(i)[v^\alpha_{n{\bf k}_\parallel}(i)]^*
}\label{gap_eq}
\end{equation}

\begin{figure}
\centerline{\includegraphics[width=0.3\textwidth]{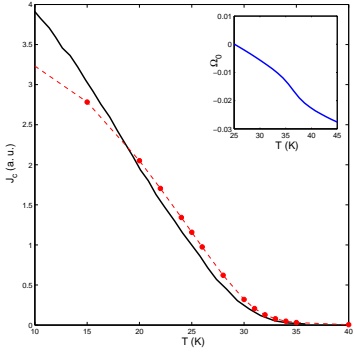}}
\caption{The critical Josephson current as a function of temperature. The parameters for the underdoped spacer are doping $x=0.1$, the magnitude of perpendicular hopping $t_c=0.05t_0$,
the magnitude of $d$-wave interaction $V=-0.54t_0$.  The black curve is for the experimental data. The inset gives the cooperon energy as a function of temperature obtained from Eq. (\ref{cooperon}) with the same parameters.}
\label{josephT}

\end{figure}

 The Josephson current between nearest-neighbor layers is given by
\begin{equation}
  I_{i,j}=(2e/\hbar)\sum_{{{\bf k}_\parallel}}t_{{{\bf k}_\parallel},ij}{\rm Im}\langle c_{{{\bf k}_\parallel}i,\uparrow}^\dagger c_{{{\bf k}_\parallel}j,\uparrow}+
 c_{{{\bf k}_\parallel}i,\downarrow}^\dagger c_{{{\bf k}_\parallel}j,\downarrow}\rangle
\end{equation}
This can be evaluated as \cite{Yeyati-PRB-95},
\begin{equation}
I_{i,j}=-{4e\over\hbar}\sum_{{\bf k}_\parallel}\int {{\rm d}\omega\over2\pi}f(\beta\omega)
{\rm Re}\left[\mathcal{G}^a_{ji}({{\bf k}_\parallel},\omega)-\mathcal{G}^r_{ji}({{\bf k}_\parallel},\omega)\right]_{11}.
\end{equation}
$I_{i,j}$ satisfies the continuity equation\cite{Fazio-PRL-95,Fazio-PRB-96} between the leads, i.e., independent of layer index. Fig.(\ref{josephT}) gives the result of the critical Josephson current as a function of temperature obtained from YRZ model. We choose the thickness of underdoped ${\rm LCO}$ $d\sim 10$ layers, the perpendicular hopping $t_c=0.05t_0$ with $t_0\sim 300meV$ \cite{Kaiyu-Andreev,YRZ}, the doping $x=0.1$, the magnitude of $d$-wave attractive interaction $V=-0.54t_0$, which gives a mean-field critical temperature around $25K$ as in the Bozovic experiments\cite{Bozovic-PRL-2004}. We find a decay length of the Josephson current around $\lambda\sim 10$(unit cell) at a  temperature  $T=30K$ which is higher than the critical temperature of the underdoped spacer $T_c\approx25K$. This longer range penetration at temperatures above the superconducting transition of the underdoped spacer is due to an enhanced proximity effect.   In contrast, for an insulating spacer without cooperon states (induced by attractive interaction), a single layer is enough to kill the Josephson coupling in our formulism. These results agree with the experiments \cite{Bozovic-PRL-2004,Bozovic-one-layer} on the stoichiometric LCO quite well.

The cooperon excitation in underdoped LSCO at temperatures $T > T_c$ with total momentum $q_z$ , appears as the pole at an energy $\Omega_{q_z}$ in the pair propagator which is defined in real space as,
\begin{widetext}
\begin{eqnarray}
G_2(d,\tau)&=&\sum_{{\bf k}_\parallel,{\bf k}_\parallel'}
d({\bf k}_\parallel)d({\bf k}'_\parallel)\nonumber
\left\langle T c_{{\bf k}_\parallel\uparrow}(0,0)c_{{-{\bf k}_\parallel\downarrow}}(0,0)
c^{\dagger}_{-{\bf k}_\parallel'\downarrow}(d,\tau)c^{\dagger}_{{\bf k}_\parallel'\uparrow}(d,\tau) \right\rangle
\end{eqnarray}
\end{widetext}

\begin{figure}
\includegraphics[width=6cm]{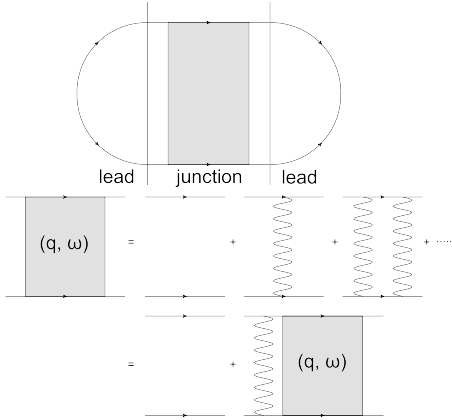}
\caption{The ladder-diagram expansion of the pair propagator and its relation to Josephson tunneling. Upper figure: Feynman diagram of Josephson tunneling; Lower figure: ladder-diagram expansion of the pair propagator. The solid line denotes the one-particle Green's function, and the wavy line denotes the effective d-wave interaction, $V'_{{\bf q}{\bf k}}=V'd({\bf k})d({\bf q})$. Note that V'= V/Z, with V the bare d-wave interaction and Z the spectral weight.  See also Schachinger and Carbotte~\cite{schachinger} }
\label{feynman}
\end{figure}

Its Fourier transform in momentum space can be estimated by summing over the particle ladder diagrams \cite{Abrikosov-textbook,Kadanoff-PR-61} See Fig. (\ref{feynman}), which gives the result,
  \begin{equation}
  G_2(0,i\omega_n)\approx{\sum_{{\bf k}_\parallel,\alpha}\left(Z^\alpha_{{\bf k}_\parallel}\right)^2X_{{\bf k}_\parallel}^\alpha(i\omega_n)d^2({\bf k}_\parallel)\over
  1+V\sum_{{\bf k}_\parallel,\alpha} d^2({\bf k}_\parallel)Z^\alpha_{{\bf k}_\parallel}X_{{\bf k}_\parallel}^\alpha(i\omega_n)}\label{G2}
  \end{equation}
when the total momentum $q_z=0$ and the perpendicular hopping $t_c$ is much smaller than the in-plane hopping $t_0$. Here the factor
$X^\alpha_{{\bf k}_\parallel}(i\omega_n)=\tanh(\beta E_{{\bf k}_\parallel}^\alpha/2)/[2E_{{\bf k}_\parallel}^\alpha+i\eta-i\omega_n]$ with $E_{{\bf k}_\parallel}^\alpha$ and $Z_{{\bf k}_\parallel}^\alpha$ denotes the dispersion and spectral weight of YRZ dispersion given in Eq.(\ref{YRZ_dispersion}). $\eta=0.01t_0$ is a small imaginary part, representing the decay of the cooperon state\cite{Rice-four-legs}.
The zero-momentum cooperon energy $\Omega_0$ is determined by the real part of the pole of pair propagator given by Eq. (\ref{G2}), i.e.,
  \begin{equation}
{\rm Re}\left[ 1+{V}\sum_{{\bf k}_\parallel,\alpha} d^2({\bf k}_\parallel)Z^\alpha_{{\bf k}\parallel}X_{{\bf k}_\parallel}^\alpha(\Omega_0)\right]=0\label{cooperon}
  \end{equation}

  The insert in Fig.(\ref{josephT}) shows the decrease (in electron notation)of $\Omega_0$ from zero as $T$ increase above $T_c$, in line with the behavior
 seen in numerical simulations \cite{Kaiyu-Troyer} of a related model.

\begin{figure}
\begin{minipage}[t]{0.2\textwidth}
\centering
\includegraphics[width=4cm]{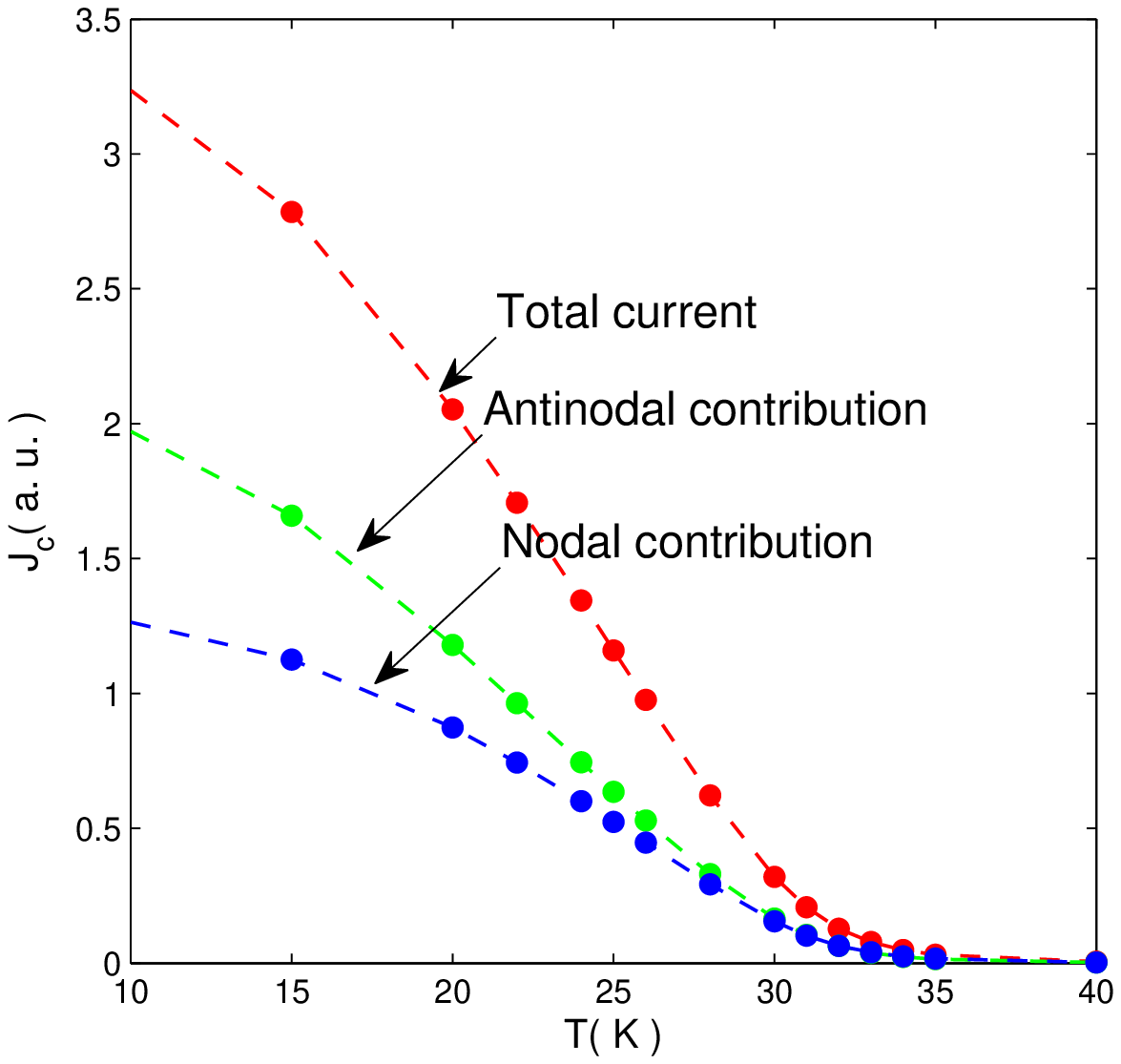}
\end{minipage}
\begin{minipage}[t]{0.2\textwidth}
\centering
\includegraphics[width=4.5cm]{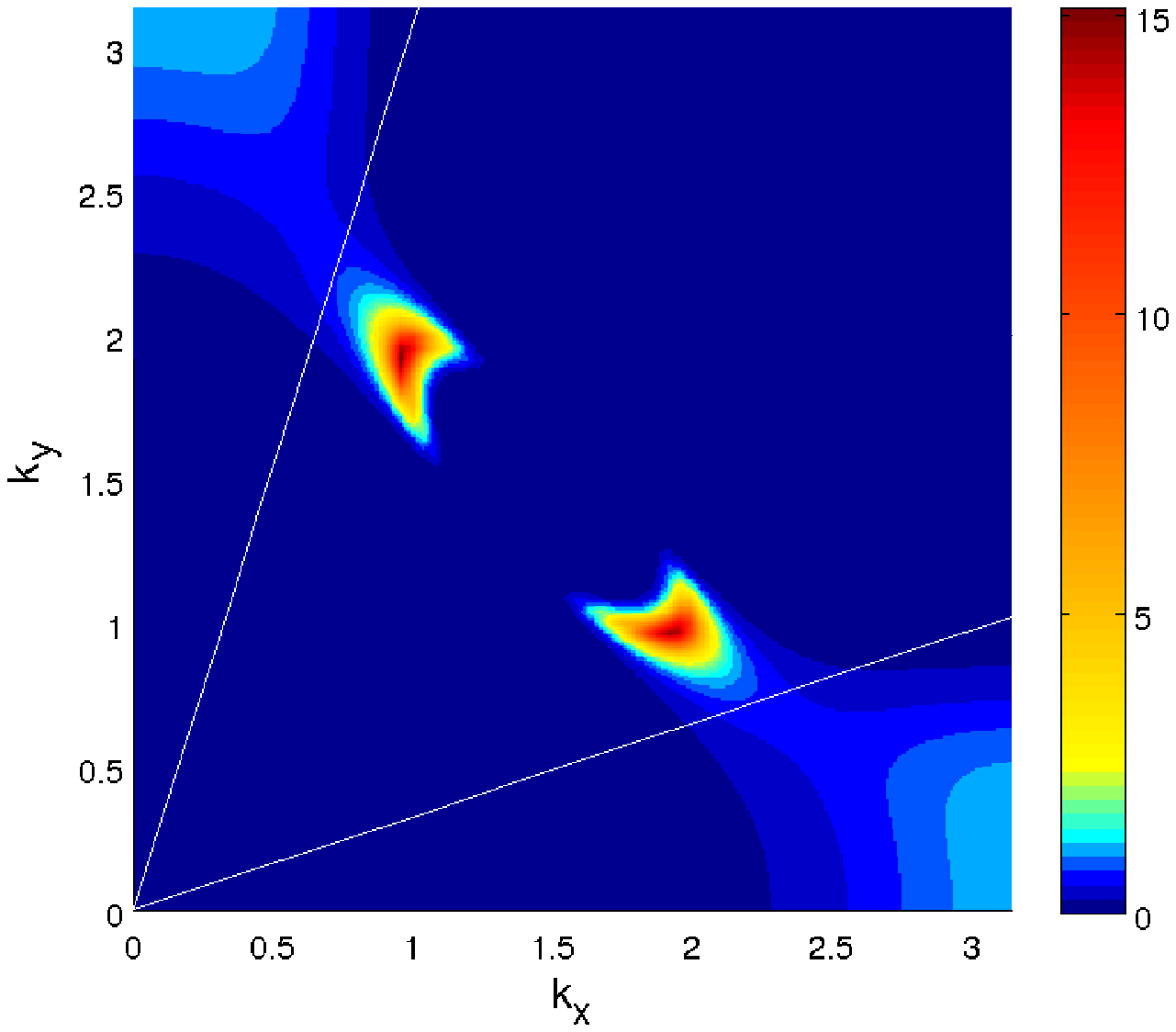}
\label{fig:side:b}
\end{minipage}
\caption{Left panel: The contribution to the Josephson current from nodal and antinodal region, respectively. Right panel: The
  distribution of Josephson current in Brillouin zone.  The choice of parameters are same as the ones in
  Fig.\ref{josephT}.  The region between the two white lines are nodal region, while the others are antinodal region.
  Tough the Josephson current is strongest at the``banana tips" of Fermi pockets, the total contribution from nodal
  region is comparable with the one from broadly spreading current around the antinodal region. }
\label{josephDis}
\end{figure}

The contribution of the cooperon excitation can be deduced from the distribution of the Josephson current in momentum
space of the underdoped sample (doping $x\approx0.1$). In the antinodal region of the Brillouin zone where cooperon
states play an important role\cite{Rice-four-legs}, the Josephson current is enhanced by cooperon states. On the other
hand, Josephson current can also be contributed from the quasi-particle excitation around the Fermi pockets centered at
nodal region of Brillouin zone\cite{YRZ}. The right panel of Fig.(\ref{josephDis}) gives the distribution of the Josephson current in
Brillouin zone, which peaks at the ``banana tips'' of the Fermi pocket.  Then we calculate the contributions from nodal
region, the region between the two white lines in the right panel of Fig.\ref{josephDis}, and the antinodal region, the region outside
the white lines.  Though the contribution at a given $\mathbf{k}$ in antinodal region is smaller than the ``banana
tips'', the relatively wide area in the antinodal region makes its total contribution to the tunneling comparable to the
contribution from the nodal region including the ``banana tips'' as shown in the left panel of fig.~\ref{josephDis}.  Note that in LSCO
the neighboring ${\rm CuO_2}$ planes along the c-axis are displaced by half a lattice parameter which reduces the
weighting of the antinodal regions\cite{Markiewicz} in $t_{\bot,ij}({{\bf k}_\parallel})$ relative to the standard interlayer hopping matrix element and the contribution of the
cooperon excitation.

There are also a few experiments on $YBCO/Y_x Pr_{1-x} Ba_2 Cu_3 O_7/YBCO$ superstructure, which also shows a very large
proximity length at $x < 0.55$\cite{Triscone-JLTP-1996,Wojek-PRB-2012}.  The spacer material YPBCO is semiconducting at
$x=0$ and only becomes superconducting with increasing Y-substitution at $x>x_c$ ( $= 0.5$).  Recently Wojek and collaborators \cite{Wojek-PRB-2012} used low-energy muon-spin rotation(LE-$\mu$SR) to measure the pentration of the superconducting order parameter into a stoichiometric insulating PBCO spacer. They reported a substantial penetration of the superconductivity with a large spacer width of 45nm. Both the contributions from
nodal quasiparticle and cooperon are weak in this case, and the strong proximity effect  needs a different explanation.

The phenomenological YRZ theory for the pseudogap phase at underdoping interprets the antinodal energy gap as a
precursor to the Mott insulator at zero doping. The analogy between the 2-dimensional and finite ladder systems has led
to proposals that a cooperon resonance appears as the antinodal energy gap opens up. The aim of this paper is to test
this proposal by examining the Josephson coupling between optimally (or overdoped) cuprate superconductors separated by
a barrier of an underdoped cuprate in its pseudogap state with a finite energy cooperon excitation which contributes to the Josephson coupling strength.  Here the calculations based on the YRZ model give a good description of the temperature dependent enhanced
Josephson tunneling.  A detailed examination of origin of the enhancement shows that the contribution from the
cooperon excitation is significant and is comparable with that from the nodal quasi-particles. 

\acknowledgments
We are grateful to C. Bernhard, R. Konik and A. Tsvelik for helpful discussions.  We acknowledge financial support in part from Hong Kong RGC GRF grant HKU706507, HKU701010, the National Natural Science
Foundation of China 10804125, and also from the Swiss Nationalfond and MANEP network. .


\begin{thebibliography}{99}
\bibitem{Triscone-JLTP-1996}
P. Fivat, J.-M. Triscone, and \O. Fischer, J. Low Temp. Phys. \textbf{105}, 1319 (1996).

\bibitem{Bozovic-PRL-2004}
I. Bozovic, G. Logvenov, M. A. J. Verhoeven, P. Caputo, E. Goldobin, and M. R. Beasley, Phys. Rev. Lett. \textbf{93}, 157002 (2004).

\bibitem{Bozovic-Meissner-effect}
E. Morenzoni, B. M. Wojek, A. Suter, T. Prokscha, G. Logvenov, and I. Bozovic, Nature (Communications) \textbf{2}, 272 (2011).

\bibitem{Wojek-PRB-2012}
B. M. Wojek \textit{et al.}, Phys. Rev. B \textbf{85}, 024505 (2012).

\bibitem{Marchand-fluctuating}
D. Marchand, L. Covaci, M. Berciu, and M. Franz, Phys. Rev. Lett. \textbf{101}, 097004 (2008).

\bibitem{Rice-four-legs}
R. M. Konik, T. M. Rice and A. M. Tsvelik, Phys. Rev. B \textbf{82}, 054501 (2010).

\bibitem{Kaiyu-Troyer}
Kai-Yu Yang, E. Kozik, Xin Wang and M. Troyer, Phys. Rev. B \textbf{83}, 214516 (2011).

\bibitem{YRZ}
Kai-Yu Yang, T. M. Rice, and Fu-Chun Zhang, Phys. Rev. B \textbf{73}, 174501 (2006).

\bibitem{Yang-nature-08}
{H.-B. Yang \textit{et al.},}
{Nature} {\textbf{456},} {77} {(2008)}.

\bibitem{Yang-EPL-09}
{Kai-Yu Yang \textit{et al.},}
{Euro Phys. Lett. } {\textbf{86},} {37002} {(2009)}.

\bibitem{Yang-PRL-11}
H.-B. Yang, J. D. Rameau, Z.-H. Pan, G. D. Gu, P. D. Johnson, H. Claus, D. G. Hinks, and T. E. Kidd, Phys. Rev. Lett. \textbf{107}, 047003 (2011).

\bibitem{AIPES-PRB-09}
{M. Hashimoto \textit{et al.},}
{Phys. Rev. B} {\textbf{79},} {140502} {(2009)}.

\bibitem{Kohsaka-Nature-08}
{Y. Kohsaka \textit{et al.},}
{Nature (London)} {\textbf{454},} {1072} {(2008)}.

\bibitem{LeBlanc-PRB-specificheat}
{ J. P. F. LeBlanc, E. J. Nicol, and J. P. Carbotte, }
{Phys. Rev. B} {\textbf{80},} {060505 (R)} {(2009)}.

\bibitem{Illes-PRB-optical}
{ E. Illes, E. J. Nicol, and J. P. Carbotte,}
{Phys. Rev. B} {\textbf{ 79},} {100505(R)} {(2009)}.

\bibitem{Carbotte-PRB-Penetration}
{J. P. Carbotte \textit{et al.},}
{Phys. Rev. B} {\textbf{81},} {014522} {(2010)}.

\bibitem{LeBlanc-PRB-Raman}
{J. P. F. LeBlanc, J. P. Carbotte, and E. J. Nicol,}
{Phys. Rev. B }{\textbf{81},} {064504} {(2010)}.

\bibitem{Leni-PRL-07}
{B. Valenzuela and E. Bascones,}
{Phys. Rev. Lett.} {\textbf{98},} {227002} {(2007)}.

\bibitem{Meng-Nature-09}
J.-Q. Meng, G. Liu, W. Zhang, L. Zhao, H. Liu, X. Jia, D. Mu, S. Liu, X. Dong, W. Lu, G. Wang, Y. Zhou, Y. Zhu, X. Wang, Z. Xu, C. Chen, and X. J. Zhou, Nature (London) \textbf{462}, 335 (2009).

\bibitem{Markiewicz}
{ R. S. Markiewicz, S.Sahrikorpi, M.Lindros, H. Lin and A. Bansil, }
{Phys. Rev. B} {\textbf{72},} {054519} {(2005)}.

\bibitem{Rice-Review}
T. M. Rice, Kai-Yu Yang, and Fu-Chun Zhang,  Rep. Prog. Phys. \textbf{75}, 016502 (2012).

\bibitem{Nozieres-EPJ-99}
P. Nozieres and F. Pistolesi, Eur. Phys. J. B \textbf{10}, 649 (1999).

\bibitem{RMFT}
F. C. Zhang, C. Gros, T. M. Rice, and H. Shiba, Supercond. Sci. Technol. \textbf{1}, 36 (1988).

\bibitem{LSR}
R. M. Konik, T. M. Rice, and A. M. Tsvelik, Phys. Rev. Lett. \textbf{96}, 086407 (2006).

\bibitem{XiangTao-PRL-96}
T. Xiang and J. M. Wheatley, Phys. Rev. Lett. \textbf{77}, 4632 (1996).

\bibitem{BdG-textbook}
P. G. de Gennes, \textit{Superconductivity of Metals and Alloys}
(Benjamin, New York, 1966).

\bibitem{Yeyati-PRB-95}
A. Levy Yeyati, A. Martin-Rodero, and F. J. Garcia-Vidal, Phys. Rev. B \textbf{51}, 3743 (1995).

\bibitem{Fazio-PRL-95}
R. Fazio, F. W. J. Hekking, and A. A. Odintsov, Phys. Rev. Lett. \textbf{74}, 1843 (1995).

\bibitem{Fazio-PRB-96}
R. Fazio, F. W. J. Hekking, and A. A. Odintsov, Phys. Rev. B \textbf{53}, 6653 (1996).

\bibitem{Kaiyu-Andreev}
Kai-Yu Yang, Kun Huang, Wei-Qiang Chen, T. M. Rice, and Fu-Chun Zhang, Phys. Rev. Lett. \textbf{105}, 167004 (2010).

\bibitem{Bozovic-one-layer}
I. Bozovic \textit{et al.}, Nature (London) \textbf{422}, 873 (2003).

\bibitem{schachinger} E. Scachinger and J. P. Carbotte, Phys. Rev. B \textbf{81}, 21451 (2010).

\bibitem{Abrikosov-textbook}
A. A. Abrikosov, L. P. Gorkov, and I. E. Dzyaloshinski, \textit{Methods of Quantum Field Theory in Statistical Physics} (Dover, New York, 1975).

\bibitem{Kadanoff-PR-61}
L. Kadanoff and P. Martin, Phys. Rev. 124, 670 (1961).

\end{thebibliography}
\end{document}